\documentclass[print]{revtex4}
\usepackage{amsmath,amssymb}
\usepackage{graphicx}

\begin{document}

\title{\large \bf  Correlated Quantum Memory: Manipulating Atomic Entanglement via Electromagnetically Induced Transparency}

\author{Hui Jing$^{a}$\footnote{Electronic address: jinghui@wipm.ac.cn}, Xiong-Jun Liu$^b$, Mo-Lin Ge$^b$ and Ming-Sheng Zhan$^a$}

\affiliation{a. State Key Laboratory of Magnetic Resonance and
Atomic and Molecular Physics,\\
 Wuhan Institute of Physics and Mathematics, CAS, Wuhan 430071, P. R. China\\
 b. Theoretical Physics Division,
Nankai Institute of Mathematics,Nankai University, Tianjin 300071,
P.R.China}

\begin{abstract}
We propose a feasible scheme of quantum state storage and
manipulation via electromagnetically induced transparency (EIT) in
flexibly $united$ multi-ensembles of three-level atoms. For
different atomic array configurations, one can properly steer the
signal and the control lights to generate different forms of
atomic entanglement within the framework of linear optics. These
results opens up new possibility for the future design of quantum
memory devices by using, e.g., an atomic grid.
  \\

\end{abstract}

\baselineskip=16pt

\maketitle

\indent The remarkable demonstration of ultraslow light speed in a
Bose-Einstein condensate in 1999 \cite{1} have stimulated rapid
advances in both experimental and theoretical works on exploring
the novel mechanism and its fascinating applications of
Electromagnetically Induced Transparency (EIT) \cite{2,3}. In
2000, Fleischhauer and Lukin presented a dark-state polaritons
(DSPs) theory, which shows an elegant mapping-and-readout quantum
memory technique by exchanging the quantum state information
between the two components of DSPs, the quantized light field and
the collective atomic excitations \cite{4,5}. The crucial
condition of adiabatic passage for the dark states was then fully
confirmed by Sun $et~al.$ by revealing the dynamical symmetry of a
single EIT medium consisting of three-level atoms \cite{6}. As an
extension, the quantum memory process in an atomic ensemble
composed of complex $N$-level $(N>3)$ atoms was also studied
\cite{7,8}, which shows more freedoms for the quantum state
control. Most recently, the storage of two entangled lights in two
$independent$ ensembles of three-level atoms was also proposed
\cite{9}, motivated by building a quantum communication network in
which the stored entanglement in each of and/or among the atomic
nodes should be conveniently manipulated \cite{10}.

In this Letter, we present a feasible technique of correlated
quantum memory via EIT mechanism in many ensembles composed of
$\Lambda$-type three-level atoms, which has build-up flexible
ability to generate and manipulate entangled states of atomic
ensembles. This differs from previous entanglement schemes (even
the EIT-based ones) \cite{5,10} since it inserts new freedoms of
manipulations between the mapping and readout processes in itself.
We resolve the general quantized DSPs model of $m$-atomic-ensemble
system and find that, for two atomic ensembles, the input probe
light can be stored in either the first or the second atomic
ensemble or both of them (in a correlated manner) by steering the
control fields. Particularly, by preparing the initial probe light
in a coherent superposition state, the entangled atomic states of
two or three ensembles can be created within the framework of
linear optics. Further manipulations of the atomic entanglement
are manifested under two configurations for three ensembles. This
scheme may have an impact on future research to design an atomic
grid as a versatile quantum memory device by, e.g., combining the
EIT and an optical lattice techniques even in a chip \cite{11}.

\noindent

\begin{figure}[ht]
\includegraphics[width=0.6\columnwidth]{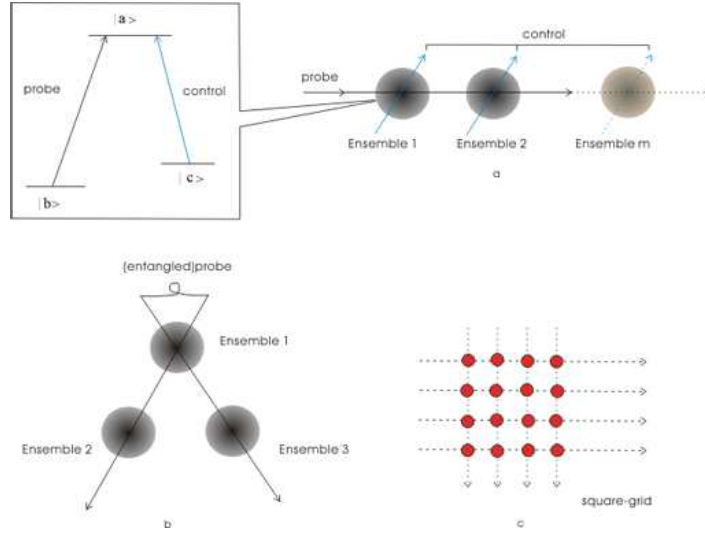}
\caption{EIT process for three or many ensembles of $\Lambda$ type
atoms located in the (a)straight-line configuration; (b)cross-line
configuration; (c)expanded square-grid configuration (the
arrow-lines represent the probe beams).} \label{}
\end{figure}

$General$ $model.-$The system we are considering is an atomic
array composed of $m$ ensembles of $\Lambda$ type
three-level-atoms (Fig.1a). Atoms of the $\sigma$-th
$(\sigma=1,2,...m)$ atomic ensemble interact with the input
single-mode quantized field with coupling constants $g_\sigma$,
and one classical control filed with time-dependent
Rabi-frequencies $\Omega_l(t)$. Generalization to multi-mode probe
pulse case is straightforward. Considering all transitions at
resonance, the interaction Hamiltonian of the system is given by
\begin{eqnarray}\label{eqn:1}
\hat V=\sum_{\sigma=1}^mg_{\sigma}\sqrt{N_{\sigma}}\hat a\hat
A_{\sigma}^{\dag}+\sum_{\sigma=1}^m\Omega_{\sigma}(t)\hat
T^+_{\sigma}+h.c.
\end{eqnarray}
where $N_{\sigma}$ denotes the atom number of the $\sigma$-th
ensemble, the collective excitation operators $ \hat
A_{\sigma}=\frac{1}{\sqrt{N_{\sigma}}}\sum_{j=1}^{N_{\sigma}}e^{-i\bf
k_{ba}\cdot\bf r^{(\sigma)}_j}\hat\sigma_{ba}^{j(\sigma)}, $ $
\hat T^{-}_{\sigma}=(\hat
T_{\sigma}^{+})^{\dagger}=\sum_{j=1}^{N_{\sigma}}e^{-i\bf
k_{ca}\cdot\bf r^{(\sigma)}_j}\hat\sigma _{ca}^{j(\sigma)}$ and
$\hat
C_{\sigma}=\frac{1}{\sqrt{N_{\sigma}}}\sum_{j=1}^{N_{\sigma}}e^{-i\bf
k_{bc}\cdot\bf r^{(\sigma)}_j}\hat\sigma_{bc}^{j(\sigma)}$; as
usual, the flip operator
$\hat\sigma^{i(\sigma)}_{\mu\nu}=|\mu\rangle_{ii}\langle\nu|
~(\mu,\nu=a,b,c)$ is defined for the $i$-th atom between states
$|\mu\rangle$ and $|\nu\rangle$ of the $\sigma$-th ensemble, and
$\bf k_{ba}$ ($\bf k_{ca}$) is the wave vectors of the quantized
(classical) optical field with $\bf k_{bc}=\bf k_{ba}-\bf k_{ca}$.
In large $N_{\sigma}$ limit and low excitation condition
\cite{6,7}, it follows that $[\hat A_{i},\hat
A^{\dag}_{j}]=\delta_{ij}, [\hat C_{i},\hat
C^{\dag}_{j}]=\delta_{ij}$ and $[\hat T^+_{i},\hat
T^-_{j}]=\delta_{ij}\hat T^z_{j}$, $[\hat T^z_{i},\hat
T^{\pm}_{j}]=\pm\delta_{ij}\hat T^{\pm}_{j}$, where $ \hat
T_{\sigma}^{z}=\sum_{j=1}^{N_{\sigma}}(e^{-i\bf k_{aa}\cdot\bf
r^{(\sigma)}_j}\hat\sigma _{aa}^{j(\sigma)}-e^{-i\bf
k_{cc}\cdot\bf r^{(\sigma)}_j}\hat\sigma _{cc}^{j(\sigma)})/2$.
Meanwhile, $[\hat T_{i}^{+},\hat A_{j}]=-\delta_{ij}\hat C_{j}$,
$[\hat T_{i}^{-},\hat C_{j}]=-\delta_{ij}\hat A_{j}$. Thereby the
dynamical symmetry of this system is governed by a semidirect
product Lie group
$(\otimes_{\sigma}su(2))\overline{\otimes}h_{2m}$, $h_{2m}$
denotes the Heisenberg algebra generated by $(\hat A_i,\hat
A_i^{\dag},\hat C_i,\hat C_i^{\dag}; i=1,2,...,m)$.

The quantum memory process of this system can be described by the
following DSPs operator
\begin{equation}\label{eqn:6}
\hat d=\cos\theta\hat a-\sin\theta\prod_{j=1}^{m-1}\cos\phi_j\hat
C_1-\sin\theta\sum_{k=2}^m\sin\phi_{k-1}\prod_{j=k}^{m-1}\cos\phi_j\hat
C_k,
\end{equation}
where the mixing angles $\theta$ and $\phi_j$ are defined through
\begin{equation}\label{eqn:7}
\tan\theta=\frac{\sqrt{g_1^2N_1\Omega_2^2\Omega_3^2...\Omega_m^2+
g_2^2N_2\Omega_1^2\Omega_3^2...\Omega_m^2+...+g_m^2N_m\Omega_1^2\Omega_2^2...\Omega_{m-1}^2}}{\Omega_1\Omega_2...\Omega_m}
\end{equation}
and
\begin{equation}\label{eqn:11}
\tan\phi_{m-1}=\frac{g_m^2\sqrt{N_m}\Omega_1\Omega_2...\Omega_{m-1}}{\sqrt{g_1^2N_1\Omega_2^2\Omega_3^2...\Omega_m^2+
g_2^2N_2\Omega_1^2\Omega_3^2...\Omega_m^2+...+g_{m-1}^2N_{m-1}\Omega_1^2\Omega_2^2...\Omega_{m-2}^2\Omega_m^2}}.
\end{equation}
Since $[\hat d,\hat d^{\dag}]=1$, $ [\hat V,\hat d \ ]=0 $, we can
obtain the general atomic dark states as
$|D_n\rangle=[n!]^{-1/2}(\hat d^{\dag})^n|0\rangle$, where
$|0\rangle=|b^{(1)}, b^{(2)},...,
b^{(m)}\rangle_{atom}\otimes|0\rangle_{photon}$
($|0\rangle_{photon}$ denotes the vacuum state of the probe field
and $|b^{(\sigma)}\rangle=|b_1,b_2,...,b_{N_{\sigma}}\rangle$ is
the collective ground state of the $\sigma$-th atomic ensemble).

In order to get more insights on the quantum state control of this
system, we first consider the special case of two atomic ensembles
and the system takes the form
\begin{eqnarray}\label{eqn:8}
|D_n\rangle=\sum^n_{k=0}\sum^{n-k}_{j=1}\sqrt{\frac{n!}{k!l!j!}}(\cos\theta)^k(-\sin\theta)^{n-k}
(\sin\phi_1)^j(\cos\phi_1)^{l}|c_{(1)}^{l},c_{(2)}^j\rangle_{spin}\otimes|k\rangle_{photon}
,
\end{eqnarray}
where $l=n-k-j$. We will show that not only the quantum memory
process still can be revealed in this quite general system but
also for the novel build-up ability to generate and manipulate the
atomic entanglement in a highly extensible style. In fact, if the
initial total state of the quantized field and atomic ensembles is
prepared in $ |\Psi_0\rangle=\sum_{n}P_n(\alpha_0)|b^{(1)},
b^{(2)}\rangle_{atom}\otimes|n\rangle_{photon} $, where
$P_n(\alpha_0)=\frac{\alpha_0^n}{\sqrt{n!}}e^{-|\alpha_0|^2/2}$ is
the probability of distribution function, then the mixing angle
$\theta$ is rotated from $0$ to $\pi/2$ while keeping the ratio
$\Omega_1/\Omega_2$ by, e.g., an acoustic-optical modulator (AOM)
\cite{12} and switching off them adiabatically
$(\tan\phi_1=\lim_{\Omega_1,
\Omega_2\rightarrow0}\bigr[g_2\sqrt{N_2}\Omega_1/g_1\sqrt{N_1}\Omega_2\bigr])$,
the total system evolves into
\begin{eqnarray}\label{eqn:10}
|\Psi_e\rangle=|\alpha_{1}\rangle_{spin1}\otimes|\alpha_{2}\rangle_{spin2}\otimes|0\rangle_{photon}
\end{eqnarray}
where $\alpha_{1}=\alpha_0\cos\phi$ and
$\alpha_{2}=\alpha_0\sin\phi$. Clearly, the injected optical state
can be converted into the atomic coherences via manipulating two
control fields. Particularly, i) if $\phi=0$, the injected light
is fully stored in the first ensemble ($\alpha_{2}=0$); ii) if
$\phi=\pi/2$, the input pulse is now stored in the second
($\alpha_{1}=0$). This mechanism can be extended to any
non-classical or entangled state of the input light.

The important issue of entanglement generation in the macroscopic
atomic ensembles hardly can be overestimated due to its practical
applications in quantum information processing \cite{11,12}. For
the present scheme, if the injected quantized field is in a
coherent superposition state, e.g., for the initial state
$|\Psi_0\rangle^{\pm}=\frac{1}{\sqrt{{\cal
N_{\pm}}(\alpha_0)}}\bigr(|\alpha_0\rangle\pm|-\alpha_0\rangle\bigr)_{photon}
\otimes|b^{(1)},b^{(2)}\rangle_{atom}$ (here ${\cal
N}_{\pm}(\alpha_0)$ is a normalized factor \cite{2}), a
two-ensemble entangled state would be created as
($|\Psi_0\rangle^{\pm}\rightarrow|\Psi_e\rangle^{\pm}$)
\begin{eqnarray}\label{eqn:entangled}
&|\Psi_e\rangle^{\pm}=\frac{1}{\sqrt{{\cal
N}_{\pm}(\alpha_0)}}|0\rangle_{photon}\otimes\bigr(|\alpha_{1},\alpha_{2}
\rangle\pm|-\alpha_{1},-\alpha_{2}\rangle\bigr)_{spin},
\end{eqnarray}
and the entanglement of atomic coherences \cite{13}
$E^{\pm}(\alpha_{1}, \alpha_{2})=-$
tr$(\rho^{\pm}_{\alpha_{1}}\ln\rho^{\pm}_{\alpha_{1}})$, with the
reduced density matrix $\rho^{\pm}_{\alpha_{1}}=$
tr$^{(\alpha_{2}, atom)}(|\Psi_e\rangle\langle\Psi_e|)^{\pm}$, can
be controlled by the two control fields. In particular, if we
start from an initial state $|\Psi_0\rangle^{-}$ and choose
$\phi=\pi/4$, we then obtain an EPR-type entangle state:
$(|+\rangle|-\rangle+|-\rangle|+\rangle)/\sqrt{2}$, where
$|\pm\rangle=\bigr(|\alpha_0/\sqrt{2}\rangle\pm|-\alpha_0/\sqrt{2}\rangle\bigr)_{spin}/\sqrt{{\cal
N}(\alpha_0/2)}$. This process may be viewed as a simple linear
optical circuit which transforms a standard basis to an entangled
one \cite{14}. Following the method developed in Ref. \cite{8} it
is straightforward also here to confirm the condition of adiabatic
evolution and therefore the the robustness of the system
\cite{7,15}. In addition, as a noticeable analogy, it deserves
further explorations by recalling our scheme of generating two
entangled $lights$ from an initial optical superposition state via
a $single$ four-state atomic medium \cite{7}.

These remarkable properties can be readily extended to the general
$m$-atomic-ensembles case. For an interesting example, we consider
the case of $m=3$ in which the dynamical symmetry is governed by
the Lie group $so(4)\otimes su(2)\overline{\otimes}h_6$ (see also
Fig.1a). Now, if the probe light is in a coherent superposition
state, e.g., $|\Psi_0\rangle^{\pm}=\frac{1}{\sqrt{{\cal
N}_{0\pm}}}\bigr(|\alpha_0\rangle\pm|\beta_0\rangle\bigr)_{photon}
\otimes|b^{(1)},b^{(2)},b^{(3)}\rangle_{atom}$ with a normalized
factor ${\cal N}_{0\pm}=2\pm2e^{-|\alpha_0-\beta_0|^2/2}$, the
achieved entangled state between three ensembles reads
($|\Psi_0\rangle^{\pm}\rightarrow|\Psi_e\rangle^{\pm}$)
\begin{eqnarray}\label{eqn:18}
&|\Psi_e\rangle^{\pm}=\frac{1}{\sqrt{{\cal
N}_{0\pm}}}|0\rangle_{photon}\otimes\bigr(|\alpha_{1},\alpha_{2},\alpha_{3}
\rangle\pm|\beta_{1},\beta_{2},\beta_{3}\rangle\bigr)_{spin}
\end{eqnarray}
where $\alpha_1=\cos\phi_1\cos\phi_2\alpha_0,
\alpha_2=\sin\phi_1\cos\phi_2\alpha_0,
\alpha_3=\sin\phi_2\alpha_0$ and
$\beta_1=\cos\phi_1\cos\phi_2\beta_0,
\beta_2=\sin\phi_1\cos\phi_2\beta_0$, $\beta_3=\sin\phi_2\beta_0$.
If we choose $\phi=\pi/4$ and $\varphi=\tan^{-1}(\sqrt{2}/2)$, we
can get a "GHZ-like" entangled state: $(|\alpha,\alpha,\alpha
\rangle\pm|\beta,\beta,\beta\rangle)_{spin}/\sqrt{{\cal
N}_{0\pm}}$, where $\alpha=\alpha_0/\sqrt{3}$, $
\beta=\beta_0/\sqrt{3}$. Notably, with the orthogonal basis
$|\pm\rangle\propto\bigr(|\alpha\rangle\pm|\beta\rangle\bigr)_{spin}$
(for a normalized factor), this state also can be put into: $
\Phi_{123}(\pm)=
\xi_\pm|\pm\rangle|\pm\rangle|\pm\rangle+\zeta_\pm|W_\pm\rangle $,
where $\xi_\pm$ and $\zeta_\pm$ are the normalized factors, and
$|W_\pm\rangle=|\pm\rangle|\mp\rangle|\mp\rangle+|\mp\rangle|\pm\rangle|\mp\rangle+|\mp\rangle|\mp\rangle|\pm\rangle$
is the $W$-state \cite{14}. It is fascinating to see that the
two-ensemble state is $still$ entangled after reducing the third
one, and the general feature of coherent entanglement oscillations
of Ramsey fringes for a highly entangled array also should be
observed \cite{12}.

$Expanded$ $illustration.-$Finally we prove that the novel
characteristics in the above scheme can be flexibly extended to
other different atomic configurations. As a concrete example, we
consider the three ensembles of $\Lambda$ type atoms with such an
atomic array (see Fig.1b): one probe light beam is injected to
interact with the atoms of the first and second ensembles with the
coupling constants $g_1$ and $g'_1$, while another beam interacts
with the first and third ones with the coupling $g_2$ and $g'_2$,
and the three classical control fields couple the transitions
$|c\rangle\rightarrow|a\rangle$ with time-dependent
Rabi-frequencies $\Omega_{\sigma}(t)$ $(\sigma=1,2,3)$ \cite{12}.
For simplicity, we still consider the single-mode probe lights;
the generalizations to the multi-mode case is straightforward. The
interaction Hamiltonian of the total system now is
\begin{eqnarray}\label{eqn:1}
\hat V&=&g_1\sqrt{N_1}\hat a_1\hat A_1^{\dag}+g'_1\sqrt{N_2}\hat
a_1\hat A_2^{\dag}+g_2\sqrt{N_1}\hat a_2\hat
A_1^{\dag}+g'_2\sqrt{N_3}\hat a_2\hat
A_3^{\dag}\nonumber\\
&&+\Omega_1(t)\hat T^+_1+\Omega_2(t)\hat T^{+}_2+\Omega_3(t)\hat
T^+_3+h.c.
\end{eqnarray}
and the operators here are in the same definitions as the above.
The dynamical symmetry of this system is governed by the Lie
algebra $so(4)\otimes su(2)\overline{\otimes}h_6$ and now the DSPs
operator is given by
\begin{equation}\label{eqn:2}
\hat d=\hat d_1 + \hat d_2,
\end{equation}
where $\hat d_1=\cos\theta_1\hat a_1-\sin\theta_1\cos\phi_1\hat
C_1-\sin\theta_1\sin\phi_1\hat C_2$, $\hat d_2=\cos\theta_2\hat
a_2-\sin\theta_2\cos\phi_2\hat C_1-\sin\theta_2\sin\phi_2\hat C_3$
and the mixing angles $\theta_{1,2}$ and $\phi_{1,2}$ are defined
just as Eq. (3)-(4), e.g.,
$\tan\phi_1=g'_1\sqrt{N_2}\Omega_1/(g_1\sqrt{N_1}\Omega_2)$ and
$\tan\phi_2=g'_2\sqrt{N_3}\Omega_1/(g_2\sqrt{N_1}\Omega_2)$. It
can be readily verified that $[\hat d, \hat d^{\dag}]=1$ and
$[\hat d, \hat V]=0$. Now we consider the initial state of the
system with two entangled probe lights:
$|\Psi_0\rangle^{\pm}=\frac{1}{\sqrt{{\cal
N}_{0\pm}}}\bigr(|\alpha_0\rangle_1|\alpha_0\rangle_2\pm|\beta_0\rangle_1|\beta_0\rangle_2\bigr)_{photon}
\otimes|b^{(1)},b^{(2)},b^{(3)}\rangle_{atom}$ (${\cal N}_{0\pm}$
is a normalized factor), then it turns out that the atomic
entangled state of the three ensembles as Eq. (8) can be obtained
again by properly steering the control fields, but with
$different$ parameters ($\eta=\alpha,~\beta$):
$\eta_1=(\cos\phi_1+\cos\phi_2)\alpha_0,
\eta_2=\sin\phi_1\alpha_0$ and $\eta_3=\sin\phi_2\alpha_0$. In
particular, if $\phi_1=\phi_2=\pi/2$ or the Rabi frequency
$\Omega_1$ is kept much smaller than both $\Omega_2$ and
$\Omega_3$ during the process that the three control fields are
turned off, we can arrive at the reduced two-ensemble entanglement
of the three atomic ensembles
\begin{eqnarray}\label{eqn:19}
\Phi_{atom}(\pm)=\frac{1}{\sqrt{{\cal
N}_{0\pm}}}|b^{(1)}\rangle\otimes\bigr(|\alpha_0\rangle_1|\alpha_0\rangle_2
\pm |\beta_0\rangle_1|\beta_0\rangle_2\bigr)_{spin}.
\end{eqnarray}
The entanglement of two probe lights are fully transferred into
that of the second and third ensembles! Clearly, this way to
create the atomic entanglement differs itself from previous
schemes \cite{5,10,11}.

In conclusion, we have proposed a feasible scheme to achieve
correlated quantum memory of photons among many atomic ensembles
composed of identical three-level atoms. The essential feature of
the present scheme is that both the storage style and the quantum
entanglement form can be controlled just using the "standard" EIT
procedure in a "single" memory node within an actual quantum
network. Our method is based on a general quantized DSPs model of
the united system of $m$-atomic-ensembles from which we find that,
for the special case of two atomic ensembles, the input quantized
light can be stored in either the first or the second ensemble or
both of them (in a correlated manner) by steering the external
control fields. In particular, by preparing an initial probe light
in a coherent superposition state through, e.g., a
beam-splitter-based catalysis technique \cite{16}, an atomic
entanglement of two ensembles can be created within the framework
of linear optics. The interesting analogy with the EIT process in
a single four-state atomic ensemble and the validity of adiabatic
passage conditions are also confirmed. The manipulations of atomic
entangled state are explored for the case of three ensembles under
two different configurations, which can be flexibly extended to
more complex structures like a tree or square array (see e.g.,
Fig.1c). This may have a strong impact on future research in
designing new versatile quantum memory devices by using an atomic
grid based on an EIT-type experiment.

\bigskip

\noindent We acknowledge Profs. J. Wang, H. Xiong and K. Gao for
their discussions. We also thanks the kind helps of Xin Liu and Y.
Zou. This work was supported by NSFC No.10275036 and No.10304020.





\end{document}